\documentclass[aps,floats,twocolumn, preprintnumbers,tightenlines]{revtex4}
\usepackage{amsmath}
\usepackage{amssymb}
\usepackage{amsthm}
\usepackage{graphics}
\usepackage{epsfig}

\newcommand{\bea}{\begin{eqnarray}}
\newcommand{\eea}{\end{eqnarray}}
\newcommand{\beq}{\begin{equation}}
\newcommand{\eeq}{\end{equation}}

\begin{document}

\title{Branes and Black holes in Collision}
\author{Antonino Flachi}
\email{flachi@yukawa.kyoto-u.ac.jp}
\affiliation{Yukawa Institute for Theoretical Physics, Kyoto University, Kyoto 606-8502, Japan}
\author{Takahiro Tanaka}
\email{tama@scphys.kyoto-u.ac.jp} 
\affiliation{Department of Physics, Kyoto University, Kyoto 606-8502, Japan}
\date{February 2007}
\preprint{YITP-07-08}
\preprint{KUNS-2064} 

\begin{abstract}
We study the collision of a brane with a black hole. Our aim is to
 explore the topology changing process of perforation of a brane. The
 brane is described as a field theoretical domain wall in the context of
 an axion-like model consisting of a complex scalar effective field
 theory with approximate $U(1)$ symmetry. We simulate numerically the
 dynamics of the collision and illustrate the transition from the
 configuration without a hole to the pierced one with the aid of a phase
 diagram. The process of perforation is found to depend on the collisional velocity, and, contrary to our expectation, we observe that above a critical value of the velocity, the black hole has no chance to perforate the wall. That is: high energy collisions do not assist piercing.
We also show that, only when the model parameters are fine-tuned 
so that the energy scale of the string is very close to that of 
the domain wall, the collision of the wall with the black hole has a possibility 
to provide a mechanism to erase domain walls, if the hole expands. 
However, in such cases, domain walls will form with many holes edged 
by a string and therefore disappear eventually. 
Therefore this mechanism is unlikely to be a solution to the cosmological domain wall problem, 
although it may cause some minor effects on the evolution of a domain wall network.
\end{abstract}
\maketitle

\section{Introduction}

A successful theory of brane collisions would provide an essential tool
to understand many features, and, on a practical level, to make specific
predictions, in the context of cosmology in string/M-theory scenarios,
where the Big Bang may be associated to the collision of a pairs of
branes. Many aspects of the problem are discussed in Ref.~\cite{gibbons}.

The theory of brane collision is being developed mainly in two
directions: a geometrical one, based on the DBI action and the theory of
minimal surfaces, and the other within a super-gravity approach, by
(exactly) solving the time-dependent equation of motion. For both
approaches see Ref.~\cite{gibbons} and the list of references given
there. The problem is also being addressed numerically within an
effective field theory context, where the branes are modelled as domain
walls (some examples are Refs.~\cite{maeda,copeland,takamizu}).  

In all the works mentioned above, the collision takes place between two
branes immersed in a higher dimensional space.
Here we would like to modify the starting point, and ask what happens
when a brane collides with another `object', that is not necessarily a
brane. The object we have in mind is a black hole.

From the point of view of the brane-universe, this is a natural question
to ask, and physical set-ups of this sort can be easily imagined at
early stages of the evolution of the universe, where perturbations might
have collapsed and formed primordial black holes. Besides, the study of
this problem can be motivated from various other directions. 
A recent one lies in the context of topology changing transition in
gravitational systems. A well studied example of this sort is provided
by the ({\it merger}) transition between a Kaluza-Klein black hole and a non-uniform 
black string \cite{gl,kolphysrept}. Such an example presents striking
similarities with the system of a brane and a black hole, as it has been
recently highlighted in Ref.~\cite{frolov2006}, where the importance of
exploring these phenomena in the context of field theory has been
stressed. The reason is that when a change in the topology of the
spacetime occurs, the curvature is expected to diverge. Thus, a field
theory model would provide a reasonable example to explore the process
of topology change in a physical situation free of singular
behaviors. Reference \cite{frolov2006} proposes a working-model to explore
these similarities: a (Dirac-Nambu-Goto) brane attached to a black hole:
a {\it brane-black hole system}. Related study has also been carried out
by the present authors both within a geometrical approach \cite{ft}, and
in the context of domain walls in field theory \cite{art1,art2}. 

In this work we wish to explore some other features of the brane-black
hole systems, which are important when field theory constructions are employed:
when the brane is described as a field theoretical domain wall, topology
changing processes depend on the specific field theory model used. Here
we will show this explicitly and consider an example where a
`topological process', which is impossible in the set-up considered in 
Refs.~\cite{frolov2006,ft,art1,art2}, can be realized: the perforation of a
brane due to the collision with a black hole and subsequent nucleation
of a string loop on the brane world-sheet. This is evidently relevant in
the context of the brane universe, in the terms described at the
beginning of this section. For this reason we consider the problem in
general dimensions and display the results of numerical simulations 
in 4D, 5D and 6D.

Let us now turn to the problem at hand: the perforation of a brane. A
similar question was investigated in flat space in
Refs.~\cite{kibble,evil1}. The main point was to show that when there is
an approximate gauge symmetry or a series of phase transitions, hybrid
topological structures may arise. One example of this sort is provided
by domain walls attached to strings, and 
concrete realizations are provided by axion models, having an exact,
spontaneously broken discrete $Z_N$ symmetry, embedded in a $U(1)$
group. The vacuum structure presents, due to the discrete symmetry, a
degeneracy, and an axion domain wall is a minimum energy configuration
which interpolates between two neighboring vacua. In some cases ($N=1$), the walls are
topologically unstable by quantum nucleation of a hole, whereas $N>1$
models are, instead, stable. Despite of this difference, the decay rate
of $N=1$ walls is exponentially suppressed. Further work, carried out in
Ref.~\cite{ce}, considered gravitational effects and has shown that, as
far as `de Sitter' walls are concerned, the nucleation of one hole is
not sufficient to completely destroy the wall, and at least four holes
are necessary for this purpose. Interestingly, as the authors of
Ref.~\cite{ce} suggest, this may signal a more generic kind of
instabilities of p-branes in supergravity models. 

In the present article, we wish to discuss the formation of a puncture
on the domain wall as a classical phenomena, occurring due to the
collision with a black hole and discuss the topological stability from a
classical standpoint. After recalling the main features of the field
theory model we employ, we will report on our results about the dynamics
of the collision. The complex scalar effective field theory of
Ref.~\cite{evil1} is used to describe the field theoretical domain
wall. Such model, in principle, allows one to consider the phenomena
of perforation. The collision between an axion domain wall and a black
hole is then simulated numerically illustrated with the aid of a phase
diagram (Fig.~\ref{crit}). 

Before jumping into the calculations we would like to comment on another
cosmological scenario for which the process we are about to study is
important: the cosmological domain wall problem. It is not our aim to
discuss it in detail (a thorough discussion and a list of references can
be found in Ref.~\cite{vilenkinbook}), rather, we would like to make some
comments on a conjectured solution, proposed in Ref.~\cite{sfs}.

It is well known that domain walls occur in any gauge theory in which a
discrete symmetry, not part of the gauge symmetry, is spontaneously
broken. Hence, it is difficult to imagine that domain
walls, regardless of what the fundamental theory may be, have not been
created in the early universe, their presence has, in most cosmological
models, disastrous consequences. This is, essentially, due to the fact
that domain walls carry too much energy and destroy the prediction of
standard Big Bang model \cite{zko}. The question is how to
avoid their domination over the energy density of the universe. 

This problem has been explored extensively, and a range of possible
solutions has been considered. Inflation is a popular way out, since it
can dilute, over an expanding volume, the energy density of the
walls. However, this mechanism may not work if the walls are created at
energies below the inflationary scale typically of order $10^{16}$
GeV. Other remedies have been suggested. One example is the
`bias mechanism', discussed in Refs.~\cite{zko,bias,kibble1}, which
introduces a small difference in the energy density of the vacua on the
two sides of the wall. As a result, the wall may collapse. This mechanism,
however, suffers from some limitations, and in realistic models,
instead of decaying, the walls may form stable bound state structures
\cite{dvalinano}. High temperature symmetry non restoration, dissipation
due to interaction of the walls with the surrounding matter, and many
other possibilities have been discussed in the literature (see
\cite{vilenkinbook} for an extensive review), but, to the best of our
knowledge, none of the possible remedies proposed so far 
seem to have a `universal' character that cures the domain wall problem
in a general and model independent way. This leaves space for discussion
of other mechanisms, and here we would like to comment on the one
proposed in Ref.~\cite{sfs}. The idea is that, interacting with
primordial black holes, the domain walls are perforated and a hole
forms. Since the process is classical, and thus may not be suppressed, the
subsequent expansion of the hole may provide, in principle, an efficient
way to get rid of the walls.

Although, at first glance, the mechanism may seem attractive, when
considering specific realizations, one can easily argue that it may only
work in cosmologically relatively safe situations. The reason is simply that a
hole, when created, must be bounded by a string. In other words, we must
consider an hybrid domain wall-string system, like those arising in
axion models, mentioned above. Evidently, the mechanism does not work in
the case of $N>1$ ($N$ essentially tells us how many `domain wall
branches' are attached to the string), since the production of a hole is
topologically impossible, {\it i.e.} in the $N>1$ case the
different domains are in different vacua. If a hole is formed, one can
go from one vacuum to another without going over a potential barrier,
which is impossible for $N>1$.
In the case of $N=1$, there is only one vacua, and the interpolating
field configuration goes around the bottom of the potential. The
meta-stability of such walls makes them harmless for cosmology, since
the walls rapidly break up as a result of multiple self-intersections and
decay into elementary particle and gravitational waves, as discussed in 
Ref.~\cite{evil1}. A recent review on axion cosmology can be found on
Ref.~\cite{sikiviereview2006}.

The previous considerations make the mechanism proposed in Ref.~\cite{sfs}
less appealing (one may also try to imagine more complicated topological
processes in the case $N>1$, however the simulation we carried out in
the case $N=2,3$ suggest the only recombination will take place). 
However, if the energy scale at which strings form is much larger than that at which domain walls form, inflation can occur in between, and usual dynamical processes will not be an efficient way of destroying the walls. It is, thus, reasonable to ask whether the collision with primordial black
holes may enhance the destruction of the wall and compete with other
mechanisms in specific examples. 

Anticipating the results we will discuss in this paper, the mechanism
turns out to be efficient only if the domain walls form at energy 
scales immediately below those at which strings form. 
However, in this case the domain walls have a lot of holes 
edged by a sting from the beginning and therefore are cosmologically 
harmless. For this reason the mechanism of perforation turns out to 
be less attractive than it might seem at first sight.

\section{Axionic branes}

Axionic domain walls were introduced after Peccei and Quinn proposed a
solution to the strong CP problem \cite{pq}, and were studied in the
context of topological defects and cosmology 
(See for example Refs.~\cite{evil1,kibble,sikivie1,linde}). 
The Peccei-Quinn symmetry is a global $U(1)$
symmetry which is broken at an energy scale $f_\theta$ typically constrained
by accelerator experiments and astrophysics, leaving the energy scale
$f_\theta$ to vary within the so called axion window, between 
$10^{10}$ GeV and $10^{12}$ GeV. This corresponds to constraining the
axion mass roughly within the range between 
$10^{-5}$ eV and $10^{-2}$ eV. Due to the
$U(1)$ symmetry, global strings are produced during the symmetry
breaking. The axion, initially massless, becomes massive at the QCD
scale due to instanton effects. When the temperature, after reheating, 
reaches values of order of $100$ MeV, the potential acquires a non vanishing term:
\beq
U_N = C m_\pi^4 \left( 1 - \cos N \theta/f_\theta \right)~,
\label{potax}
\eeq
with $\theta$ being the axion field, $C$ a constant of $O(1)$ and $N$
an integer which depends on the detailed structure of the model. The
above potential has a $Z_N$ symmetry which accounts for domain wall
solutions. 

Here we will consider a $(p+2)-$dimensional effective theory described by the following action:
\beq
S=\int d^{p+2}x ~L~,
\eeq
where
\beq
L = - \partial_\mu \Phi^\dagger \partial^\mu \Phi -U(\Phi, \Phi^\dagger)~,
\label{aN}
\eeq
where 
\beq
U(\Phi, \Phi^\dagger)={\lambda\over 4}  \left(\Phi \Phi^\dagger - \eta^2\right)^2-2\mu^2\Phi^2(\cos \theta - 1)~.
\label{potential_mod}
\eeq
The above effective theory, corresponding to the axion model with $N=1$,
will be our working model. The complex scalar field $\Phi$ is
parametrised as $\Phi= \rho~ e^{i\theta}$ and $\theta$ represents the axion
field. This model has a unique vacuum state at $|\Phi| = \eta$, $\theta
= 0$ and no exact symmetry. However 
the model has an approximate $U(1)$ symmetry for $\mu^2 \ll \lambda
\eta^2$, and there are accompanying global string solutions. 
If the symmetry were
exact, the phase $\theta$ would change uniformly around the string, but
the second term in the potential forces $\theta$ to zero, so that all
the variation of $\theta$ from $0$ to $2\pi$ is confined to a domain
wall centered at $\theta = \pi$, whose boundary is attached to a string. The
thickness of the domain wall $\sim\mu^{-1}$ 
is much larger than that of the string core 
$\sim(\sqrt{\lambda}\eta)^{-1}$. 
To see the presence of domain wall solutions, it is sufficient to notice
that away from the string core, the VEV of $\Phi$ is $<\Phi> = \eta
e^{i\theta}$, and the above Lagrangian becomes
$$
L = \left(
-\partial_\mu \theta^\dagger \partial^\mu \theta -2\mu^2(\cos \theta -1)\right)\eta^2
~,
$$
which is of the form of a sine-Gordon model. In such a case, plane-symmetric domain
wall solutions are well known to be of the form
$$
\theta(x) = 4 \arctan e^{\mu z}~,
$$
where $z$ is a coordinate perpendicular to the wall. 

We record here some useful scaling properties of the above $N=1$ axion
model. Notice that upon a coordinate dilatation,
$$
x\rightarrow x/a~,
$$
followed by 
\bea
\Phi &\rightarrow&  \Phi/\eta_0\nonumber \\
\eta &\rightarrow& \eta/\eta_0 \nonumber 
\eea
and
\bea
\lambda &\rightarrow& a^2 \eta_0^2\lambda \nonumber \\
\mu^2 &\rightarrow& a^2 \mu^2 \nonumber
\eea
the action scales as
\bea
S \rightarrow  {a^{-p}\over \eta_0^2}S.
\eea
It is trivial to see that: we are free to rescale $\eta$ to
unity (in four dimensions, energies will be measured, then, in fractions
of $\eta$); and that a coordinate dilatation at the classical level corresponds to simultaneous variation of the coefficients $\lambda$ and $\mu^2$.

We conclude the section by commenting on the generality of the model considered here.
It can be easily understood that for the process of perforation to occur, the essential feature is that the brane must be metastable, meaning that the formation of a hole bounded by a string has to be allowed. Therefore the potential must be characterized by at least three quantities: one giving the energy scale at which the domain wall forms; the second being the size of the vacuum manifold that {\it accounts} for string formation; and, finally, the height of the potential {\it leading} to string formation. All these features are comprised in the above model (\ref{potential_mod}), which depends on three (free) parameters ($\eta$, $\lambda$ and $\mu$) that completely characterize the above three quantities.

The above scaling properties turns out to be a convenient bonus when considering the interaction between a black hole and the domain wall.
Defining the ratio of the portion of the wall accreted by the black hole and the mass of the black hole as
$$
\epsilon \equiv {\sigma_w R^{p}\over R^{p-1}} = \mu \eta^2 R~,
$$
the probe brane approximation can be written as 
$$
\epsilon \ll 1~.
$$
Here $R$ is a typical scale of order of the black hole horizon and $\sigma_w\approx \mu \eta^2$ is the tension of the wall. Now, in view of the above scaling properties, the quantity $\epsilon$ changes as 
$$
\epsilon \rightarrow \epsilon/\eta_0^2~.
$$
It is then immediate to understand that the above scaling essentially enforces the probe brane approximation.

It is also easy to see that in a cosmological context the probe brane approximation is valid. 
If we assume that the wall do not dominate the energy density of the universe, then the following inequality follows (from the Friedmann equation):
$$
\sigma_w H^{-1} < 1~.
$$ 
Assuming that the size black hole is much smaller the Hubble radius, one recovers the relation $\epsilon \ll 1$.
Having said this, we can now safely move on to the next section.

\section{collisions with black holes}

Axionic branes are metastable, in the sense that a string loop can be
nucleated spontaneously and, in principle, can destroy, at least
locally, the wall. However, since the nucleation rate is exponentially
suppressed (and even if a hole is created, the brane may inflate with a
higher expansion rate than the hole), it is not clear whether the brane
survives on cosmological time-scales. For this reason, we want to
investigate a classical (and thus not suppressed) mechanism of hole
nucleation, as the collision with a primordial black hole may be. This
process is described by the Klein-Gordon equation for the complex scalar
field $\Phi$,
\bea
%-\nabla_\mu\nabla^\mu 
%\Phi^* + {\partial U \over \partial \Phi} &=& 0~, \nonumber\\
-\nabla_\mu\nabla^\mu 
\Phi + {\partial U \over \partial \Phi^*} &=& 0~. \nonumber
\eea
We consider the spacetime of a $(p+2)$-dimensional black hole,
\beq
ds^2 = -f dt^2 + f^{-1} dr^2 + r^2 d\Omega_{p}^2~,
\nonumber
\eeq
with
\beq
 f\equiv 1-{1\over r^{p-1}}. 
\eeq
Here we could set the horizon radius $r_H$ to unity without loss of 
generality because the scaling property mentioned above. 
Then the Klein-Gordon equation takes the form
\bea
0=&-&f^{-1} \ddot{\psi_\pm} + {1\over r^{p}} \partial_r(r^{p} f \partial_r \psi_\pm) \nonumber\\
&+& {1\over r^2 \sin^{p-1}\theta} \partial_\theta (\sin^{p-1}\theta\partial_\theta \psi_\pm)  
+ {\mathcal U}_\pm~,
 \label{pde}
%0&=&-f^{-1} \ddot{\sigma} + {1\over r^{d-1}} \partial_r(r^{d-1} f \partial_r \sigma) \nonumber
%&+& {1\over r^2 \sin^{d-2}\theta} \partial_\theta (\sin^{d-2}\theta\partial_\theta \sigma) + {1\over 2i} \left( {\partial U \over \partial \Phi} - {\partial U \over \partial \Phi^*} \right) 
\eea
where we have defined $\Phi \equiv \psi_+ + i \psi_-$ and 
%\bea
%{\mathcal U}_+ &=& {1\over 2} \left({\partial U \over \partial \Phi^*}+{\partial U \over \partial \Phi} \right)\nonumber \\
%&=& {1\over 2} \lambda(\psi_+^2 +\psi_-^2 - \eta^2)\psi_+ - {\eta^2 \mu^2}
%{\psi_-^2 \over (\psi_+^2+\psi_-^2)^{3/2}}~,\nonumber \\
%{\mathcal U}_- &=& {1\over 2i} \left( {\partial U \over \partial \Phi^*} - {\partial U \over \partial \Phi} \right)\nonumber \\
%&=& {1\over 2} \lambda(\psi_+^2 +\psi_-^2 - \eta^2) \psi_- + {\eta^2 \mu^2}{\psi_+ \psi_- \over (\psi_+^2 + \psi_-^2)^{3/2}}~.\nonumber 
%\eea
\bea
{\mathcal U}_+ -i {\mathcal U}_- &\equiv & {\partial U \over \partial
\Phi}. 
\eea
The collision process, in the probe brane approximation, is then
described by the above time-dependent system of partial differential
equations with appropriate boundary and initial conditions. 

Initially the domain wall is located at a distance $z_0$ from the
equatorial plane of the black hole. The initial configuration
can be obtained by using, for example, the
relaxation method. This solution will not differ much, as intuition
suggests, from the static solution in flat space if $z_0$ is large
enough, {\it i.e.}, if the brane is far away from the black hole. And, in
fact, the relaxed solution, can be compared with the `flat space'
configuration in the limit $\mu^2 \ll \lambda \eta^2$:
\bea
\rho(t=0,r,\theta) = \eta~,~~\theta (t=0,r,\theta) = 4 \arctan e^{\mu(z-z_0)}~.\nonumber
\eea 
Here $z=r\cos \theta$. The motion of the domain wall at the initial
stages of its evolution (still far away from the black hole) is
described by boosting the solution 
$$
\psi_\pm^{(v)}(z,t)= \psi_\pm^{(0)}(\gamma\left({(z-z_0)-vt}\right))~,
$$ 
where $\gamma\equiv 1/\sqrt{1-v^2}$ 
is the Lorentz factor. The previous expression will describe the initial
conditions.
In order to determine the evolution of the system we also need to fix
the boundary conditions, that will describe the motion of the black hole
during the collision far away from the hole. For the numerical
simulation `far away' means several times the horizon size. 
In the numerical
simulation we will take a relaxed flat space solution $\psi_\pm^{(v)}(z,t)$ 
as a configuration to describe the initial and infinity 
boundary conditions.  
At the horizon the boundary conditions are written as 
$$
\partial_t \psi_\pm(t, r_H, \theta) = 0~, 
$$
since the lapse vanishes there in our coordinates. 

To study the evolution of the system we have to solve the system of PDE
(\ref{pde}). To impose the horizon boundary conditions, spherical
coordinates are appropriate. Then we need to treat near the $z$-axis 
a little carefully.  
A convenient way to handle this problem is to use the harmonic
decomposition for the angular coordinate. 
We initially decompose the solution in
terms of a complete basis of smooth global functions in the $\theta$
direction: 
\bea
\psi_\pm(t,r,\theta) &=& \sum_{k=0}^n u^{\pm}_k(t,r) \phi_k(\cos
\theta), \nonumber
\eea
where $n$ is the number of harmonics used. 
We choose the basis
functions $\phi_k(\cos \theta)$ to be the Chebyshev polynomials of 
the first kind, which are regular at $\theta=0,~\pi$ and 
satisfy the following orthogonality relation 
($x=\cos \theta$):
$$
\int_{-1}^{+1} \phi_i(x) \phi_j(x) w(x) dx = {\pi\over 2} \delta_{ij}~,
$$ 
with $w(x) = 1/\sqrt{1-x^2}$. (Notice that the normalization for $\phi_0$
differs from the conventional one.) 
Using the previous decomposition, it is possible to rewrite the
evolution equations (\ref{pde}) as follows:
\beq
{\ddot{u}^{\pm}_j} = {f \over r^{p}} \partial_r\left(r^{p} 
 f \partial_r u^{\pm}_j \right) + 
   {2f\over \pi r^2} {\mathbb J}_{jk} \cdot 
u^{\pm}_k - {2f\over \pi} {\mathbb P}^{\pm}_j~,
\label{eqom}
\eeq
where 
\beq
{\mathbb J}_{ij}  = \int_{-1}^{+1} w(x) \left[ \partial^2_\theta
\phi_i(x) + (p-1)\cot\theta \partial_\theta \phi_i(x) \right]  \phi_j(x)
dx~,\nonumber
\eeq
and 
\beq
{\mathbb P}^{\pm}_j = \int_{-1}^{+1} {\mathcal U}_\pm \phi_j(x) w(x)
dx~.\nonumber
\eeq
The last step of the numerical procedure that we adopt consists of 
solving the previous equation (\ref{eqom}) by means of standard finite difference methods. 
As a check on our numerical scheme we have evolved the domain wall in flat space, modifying appropriately the boundary condition at the horizon, making sure that the wall propagates smoothly, as it must.

\section{discussion of the numerical results}

\begin{figure}[th]
\scalebox{0.8} {\includegraphics{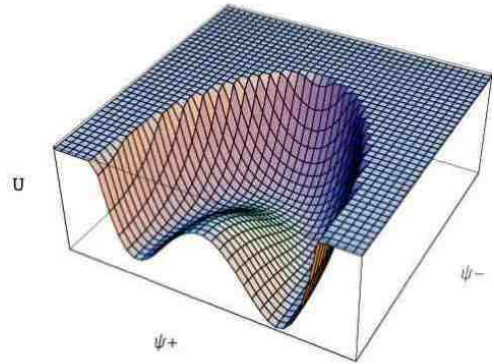}}
\caption{Potential in the $N=1$ case.}
\label{potential}
\end{figure}

Before turning to the numerical results, it is useful to discuss what is
the physically relevant range of parameters. 
First of all, it can be easily argued that for values of $\lambda$ below
a critical value, domain wall solutions cease to exist. To see this, it
is sufficient to notice that the potential has an approximately $U(1)$
symmetric string of vacua. 
Along the trajectory defined by $\Re (\Phi)=\varphi$ and $\Im (\Phi)=0$
(see Fig.~\ref{potential}), the potential assumes the form
$$
U={\lambda\over 4}(\varphi^2-\eta^2)^2+4\mu^2\varphi^2\theta(\varphi)~,
$$
where $\theta(x)$ is the step function. The previous function has two
minima, one at $\phi_T=-\eta$, the `true' one, and one at
$\phi_F=\sqrt{\eta^2-8{\mu^2\over \lambda}}$, the `false' one. Roughly
speaking the domain wall configuration interpolates between these two
vacua. It is then easy to show that the minima at $\phi_F$ becomes a
saddle point for 
$\lambda \simeq 8\mu^2/\eta^2$. 
In this case the approximate $U(1)$ symmetry is absent 
and it is obvious that domain wall solutions do not exist. 
Once the values of $\eta$ and $\mu$ are
fixed, $8\mu^2/\eta^2$ gives an upper bound on the critical value $\lambda_1$ for
$\lambda$ below which there is no static stable domain wall solution 
in flat spacetime.   

Here it is worth discussing an alternative way to estimate the 
critical value $\lambda_1$. Let us first focus on the flat space case. 
When we consider a configuration with a hole bounded by a string, 
the hole may expand or collapse, depending on its size.
If there is a stable domain wall configuration without a hole, we can argue that, 
if a hole smaller than a critical size forms, then it will collapse.
On the other hand, if the hole forms with a size larger than the critical one, 
it will expand. Hence, between these two possibilities, there must be a marginal 
(unstable) configuration at which the hole neither expands nor collapses. 
Conversely, if there is no unstable static configuration with a hole 
bounded by a string, the domain wall is connected to the expanding hole
without a barrier. In this case the domain wall configuration without a
hole will not be stable.

To see if there exists an unstable static configuration, we make use of  
the thin wall (Dirac-Nambu-Goto) approximation. 
In this approximation, we will find that this critical 
configuration always exists. 
However, once we take into account the finite thickness of the string, 
such a solution will be expected to vanish when the sting 
thickness becomes as large as the radius of the hole. 
The thickness of the string $r_s$ can be estimated from the curvature of
the scalar field potential as $r_s\approx \alpha/\sqrt{\lambda}\eta$, 
where $\alpha$ is a constant of $O(1)$.   
To evaluate the radius of the hole for the critical configuration, 
we extremize the energy of the the static configurations 
consisting of a pierced brane and a string bounding the hole 
with radius $r$, 
\begin{equation}
E\propto \mu_s r^{p-1} +\sigma_w \int_r^{\infty} dr r^{p-1},  
\label{energy}
\end{equation}
where $\mu_s\approx \eta^2$ and 
$\sigma_w\approx \mu\eta^2$ are the string and the wall tensions, 
respectively. 
The variation of the energy gives the radius of the hole at the 
critical configuration as 
\begin{equation}
r_{hole} = r_0\equiv (p-1){\mu_s\over \sigma_w}
  \approx {\beta\over \mu}, 
\end{equation} 
where we have introduced a constant $\beta$ of $O(1)$. 
Let us assume that the condition for 
a critical configuration to exist, after taking into account
the string thickness, is given by $r_s<\kappa r_{hole}$ with a constant 
$\kappa$ of $O(1)$. 
Then, the threshold value for the stability of an un-pierced 
wall is estimated as 
\begin{equation}
 \lambda_1={\alpha^2\over \beta^2\kappa^2}{\mu^2\over \eta^2}. 
\end{equation}
The dependence on $\mu$ and $\eta$ is the same as before. 
$r_0$ marks a critical values, above which a hole will expand and below which it collapses.

The above analysis has the advantage that it can be easily extended to the 
black hole background, which we will now consider taking the domain 
wall on the equatorial plane of the black hole.

Even if the scales at which strings and domain walls form are very different, and usual dynamical mechanisms do not help to destroy the walls, collisions with black holes may still have a chance to be an effective mechanism, because, when the horizon radius of the black hole is larger than the size of the critical hole, $r_0\approx \beta/\mu$, 
the domain wall will be perforated. Cosmologically we can expect this to be reasonable if formation of primordial black holes is delayed and occurs at a low scale. 
This suggests that whether the domain wall is pierced or not seems to be more or less independent of $\lambda$. However, this naive expectation is wrong. 
To show this, we repeat the analysis that we have done above 
under the presence of a black hole. 
The expression for the energy (\ref{energy})
is to be replaced with 
\begin{eqnarray*}
E\propto \mu_s f(r) r^{p-1} +\sigma_w \int_r^{\infty} dr r^{p-1},  
\end{eqnarray*}
By varying this expression with respect to $r$, 
we find
$$
0=1-{1\over 2}r^{1-p} +  \sqrt{1-{1\over r^{p-1}}}{r\over r_0}~.
$$
The coordinate radius of the hole $r_{hole}$ is determined by 
solving this equation. 

\begin{figure}[th]
\scalebox{0.8} {\includegraphics{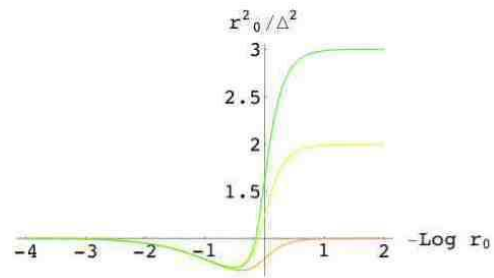}}
\caption{$r_0^2/\Delta^2$ vs $\log r_0$. 
For small $r_0$, increasing values of the dimension, increases the ratio
 $r_0^2/\Delta^2$. The curves refer (from bottom to top) to $D=4,~5,~6$ spacetime
 dimensions.}
\label{lasula}
\end{figure}

However, rather than $r_{hole}$, a more appropriate quantity to be compared 
with the string thickness $r_s$ is the proper distance between the rim of the hole and the horizon, 
$$
\Delta(r_0) = 
 \int_{1}^{r_{hole}(r_0)} f^{-1/2}(r)dr ~.
$$
We can reasonably imagine that, once the string thickness $r_s$
becomes larger than $\Delta$, 
the static configuration of a brane with a floating
string will not exist. 
Then, following the same discussion as in the case of the flat 
background, domain wall configurations with a hole of any size 
will become unstable. 
Comparing $r_s$ with $\Delta(r_0)$, it is
easy to see that this occurs for 
\begin{eqnarray*}
 \lambda > \lambda_2 \approx 
  {\alpha^2\over\beta^2\kappa^2}
   {\mu^2\over \eta^2}{r_0^2\over \Delta^2(r_0)}. 
\end{eqnarray*}
This tells us that there exists a threshold value $\lambda_2$, 
above which the black hole cannot pierce the wall.
We find that $\lambda_2$ differs from $\lambda_1$ by the factor 
$r_0^2/\Delta^2(r_0)$, which is plotted 
in Fig.~\ref{lasula} against $\log r_0$. 
Independent of the value of $r_0$, this ratio does not deviate 
much from unity. When $r_0\gg 1$ 
(recall that the horizon 
radius is normalized to unity), the existence of a black hole 
can be neglected. Then the ratio becomes unity and we have 
$\lambda_1\approx \lambda_2$. 
On the other hand, in the limit $r_0\ll 1$, 
we have $\lambda_2 \approx (p-1) \lambda_1$. 

In summary, according to the above arguments, we can expect that for
$\lambda<\lambda_1$, there will be no static domain wall configuration;
for $\lambda>\lambda_2$ the domain wall configuration exists and cannot
be perforated in the adiabatic process. 
In the intermediate range $\lambda_1<\lambda<\lambda_2$
the domain wall exists at a distance from the black hole 
and can be perforated eventually as it gets closer to the black hole.

It is now time to present the numerical results. How our 
simulation goes is summarised in Figs.~\ref{fig1-1}-\ref{fig3-2}, where
various snapshots of the evolution of the domain wall during the
collision are illustrated. We present both cases where a hole is created
and expands or only reconnection takes place. The figures show the
evolution of the field components and of the energy density, and, in
some case, to emphasize the formation of the hole, we have presented the
plots of the energy density from two different viewpoints.
The parameters of the simulation have been clipped to $n=40$ harmonics,
$400$ grid points and a maximal radial distance up to $20$ times the
horizon size. To scan the parameter space we fixed the value of the
symmetry breaking scale $\eta$ to unity and 
varied $\mu$ and $\lambda$. The
collisional velocity has been sampled from $0.9$ down to $0.001$ in units of the velocity of light.

\begin{figure}[th]
\scalebox{0.9} {\includegraphics{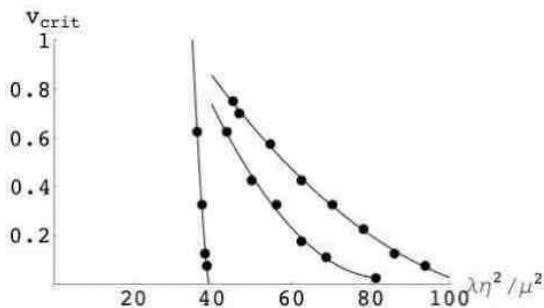}}
\caption{This figure illustrate the dependence of the critical velocity,
 above which a hole does not form, on the ratio $\lambda\eta^2/\mu^2$. It
 refers to $D=5$ dimensions and the curves, from left to right
 correspond to the small, intermediate and large $\mu$ range
 ($\mu=0.2,~0.4~,0.8$ respectively). The points are computed from the
 numerical analysis by fixing the values of $\mu$, and performing the
 simulation varying $\lambda$ and the collisional velocity.}
\label{crit}
\end{figure}

\begin{figure}[th]
\scalebox{0.4} {\includegraphics{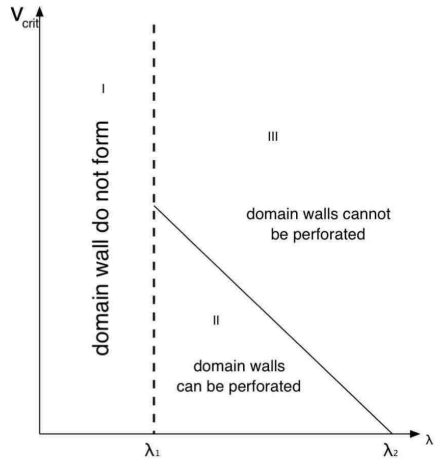}}
\caption{Schematic phase diagram representing the dependence of the critical velocity from $\lambda$. }
\label{crit_sch}
\end{figure}

\begin{figure*}[th]
\scalebox{0.8}{\includegraphics[]{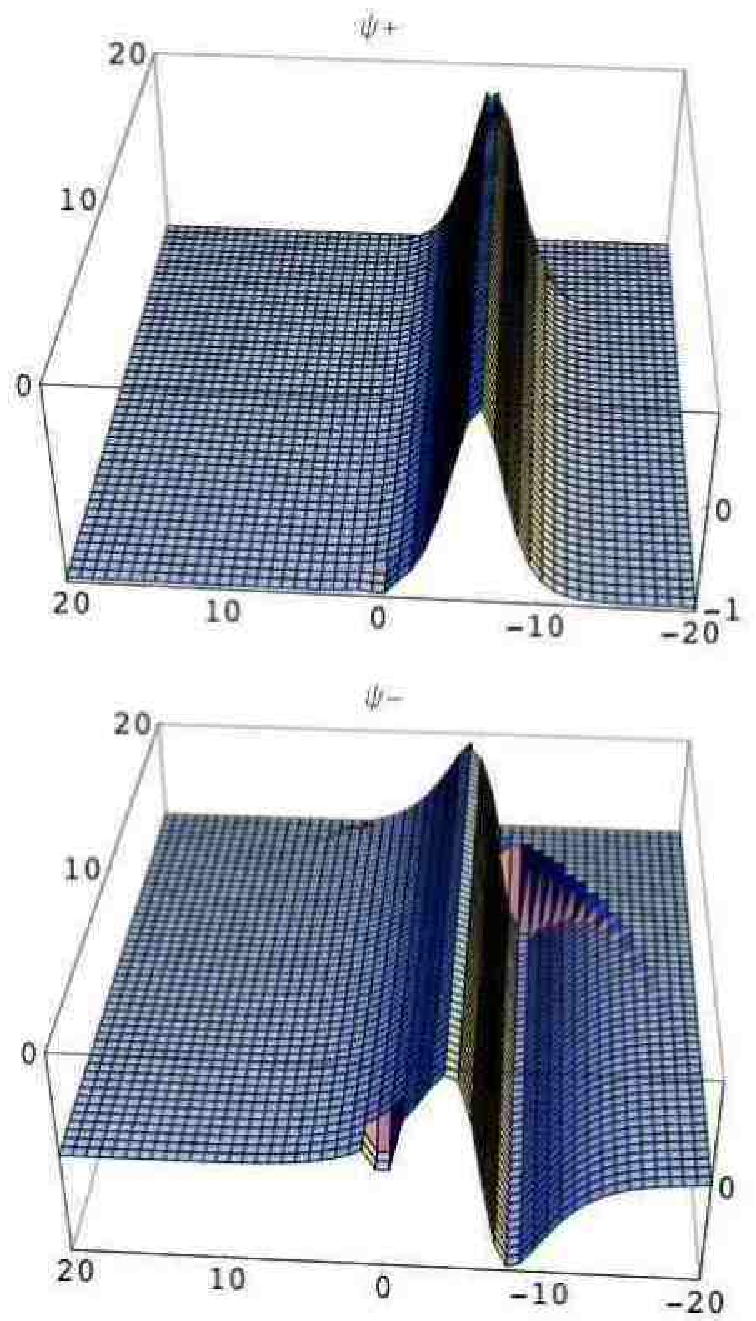}}
%\scalebox{0.6}{\includegraphics[]{fh2.eps}}
%\scalebox{0.5}{\includegraphics[]{fh3.eps}}
%\scalebox{0.6}{\includegraphics[]{fh4.eps}}
\scalebox{0.8}{\includegraphics[]{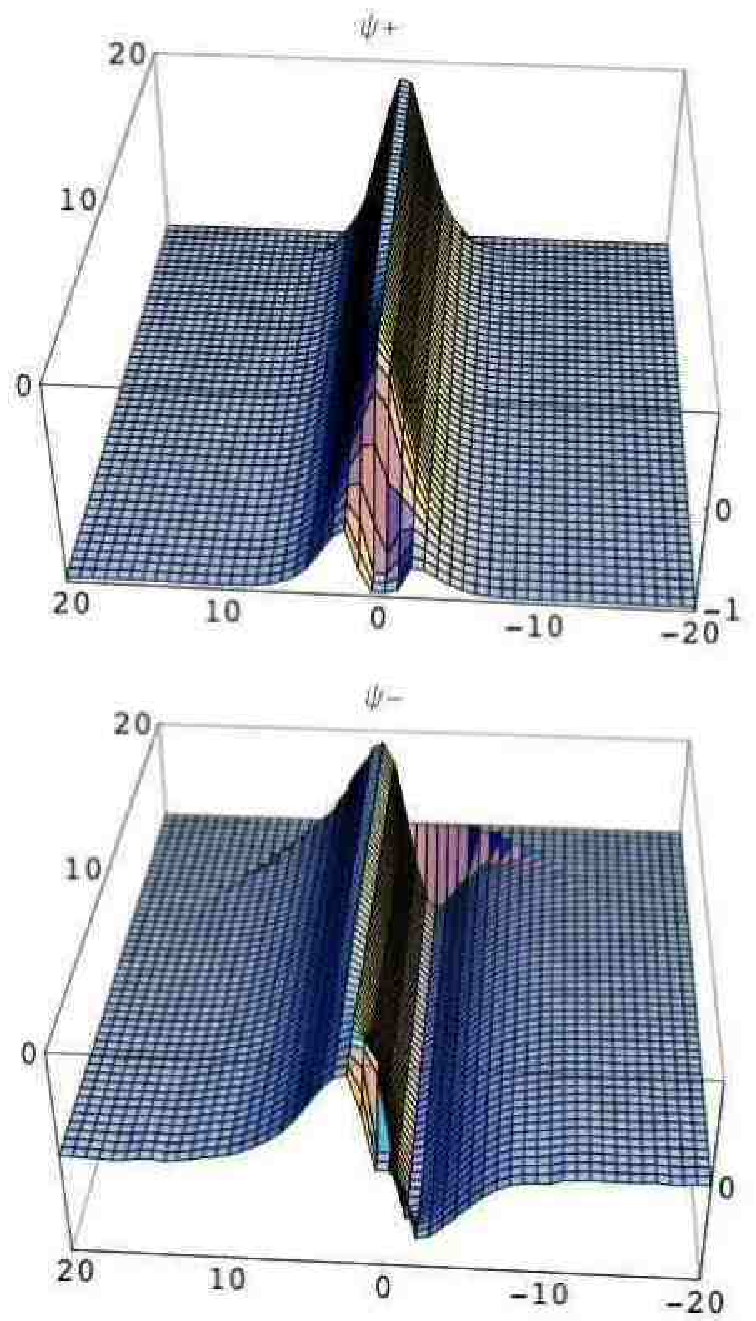}}
\scalebox{0.8}{\includegraphics[]{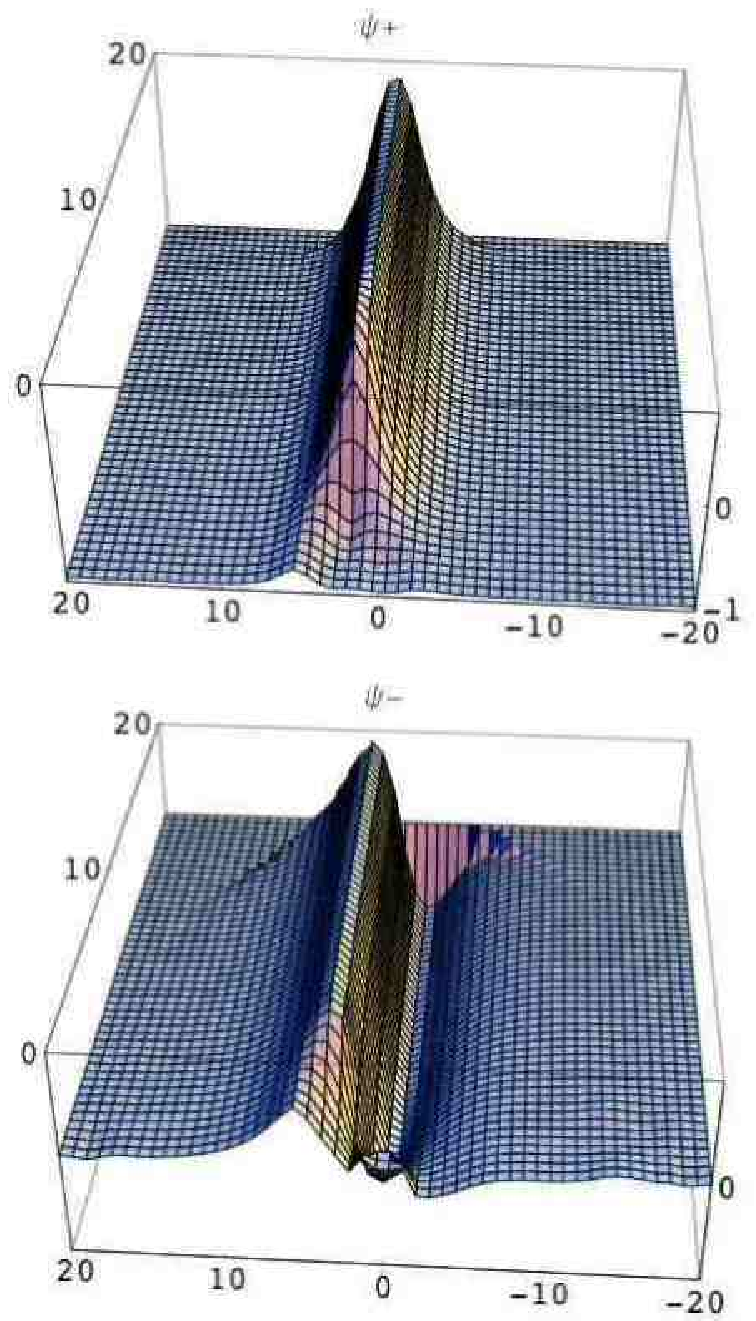}}
%\scalebox{0.5}{\includegraphics[]{fh7.eps}}
\scalebox{0.8}{\includegraphics[]{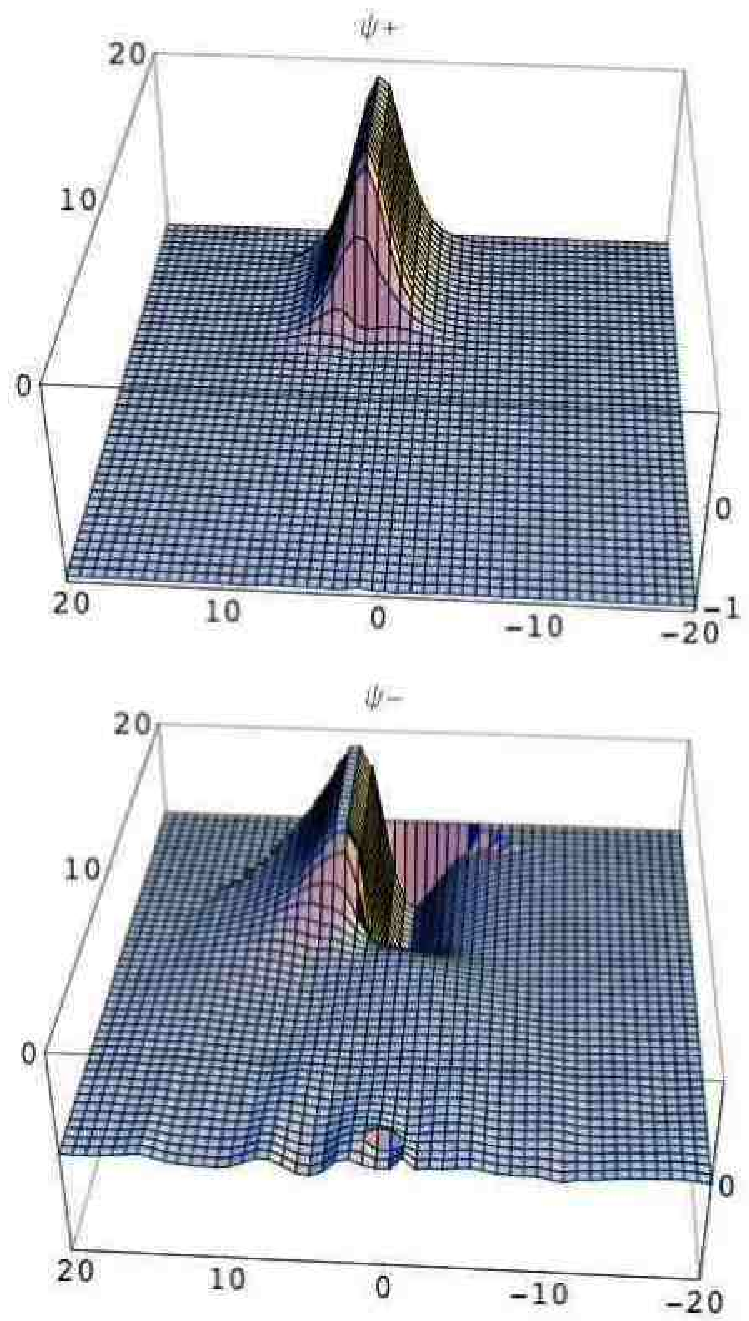}}
\caption{Snapshots of the evolution of $\psi_\pm$. The values used in
 this simulation are $\lambda=10$, $\mu=0.4$ and $\eta=1$.The velocity
 if chosen to be $0.1$. The black hole horizon is a circle of radius 1
 centered at the 0. The spacetime dimensionality is 5. One can observe
 the formation of a hole that locally destroys the wall.}
\label{fig1-1}
\end{figure*}

\begin{figure*}[th]
\scalebox{0.8} {\includegraphics[]{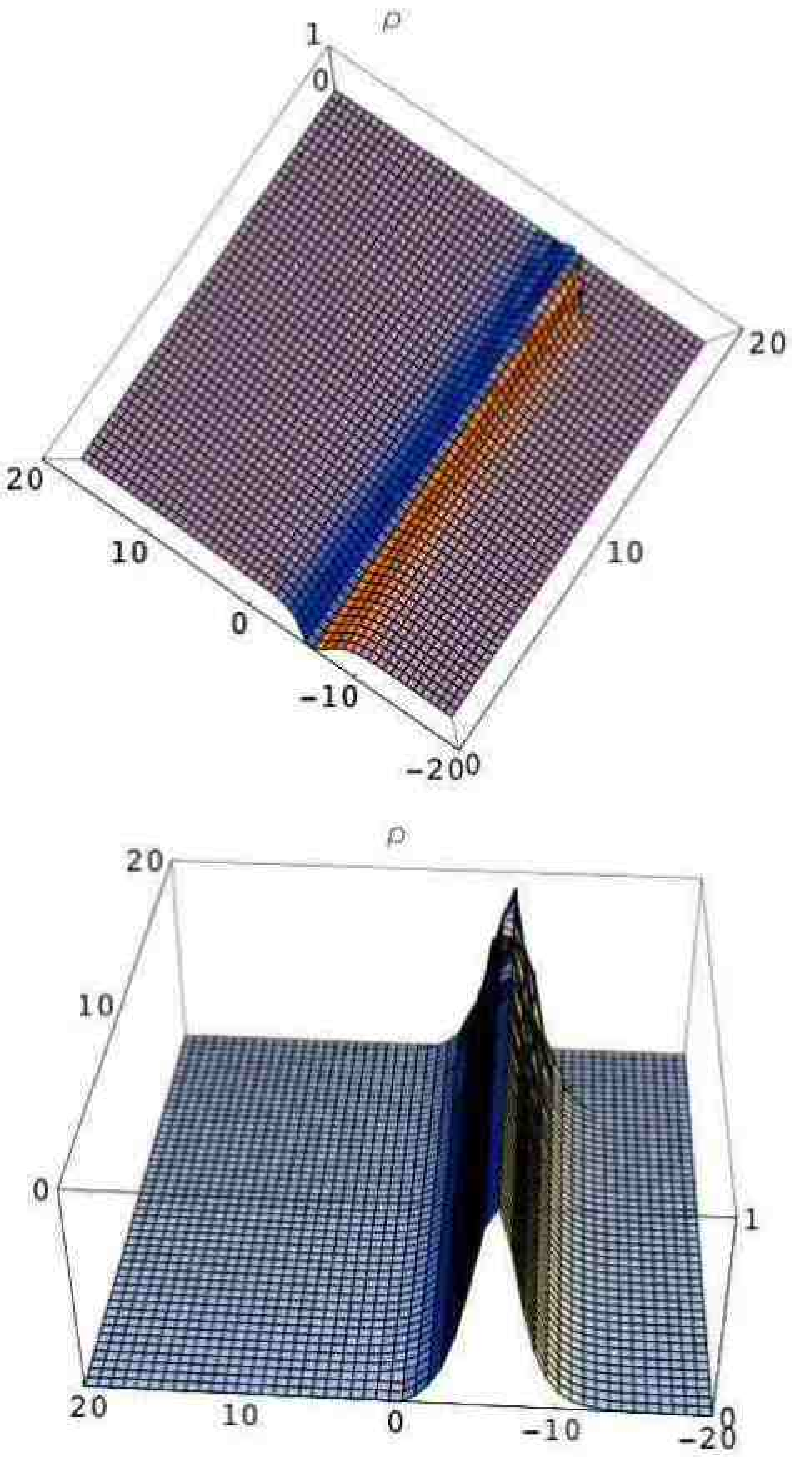}}
%\scalebox{0.6} {\includegraphics[]{edh2.eps}}
%\scalebox{0.5} {\includegraphics[]{edh3.eps}}
%\scalebox{0.6} {\includegraphics[]{edh4.eps}}
\scalebox{0.8} {\includegraphics[]{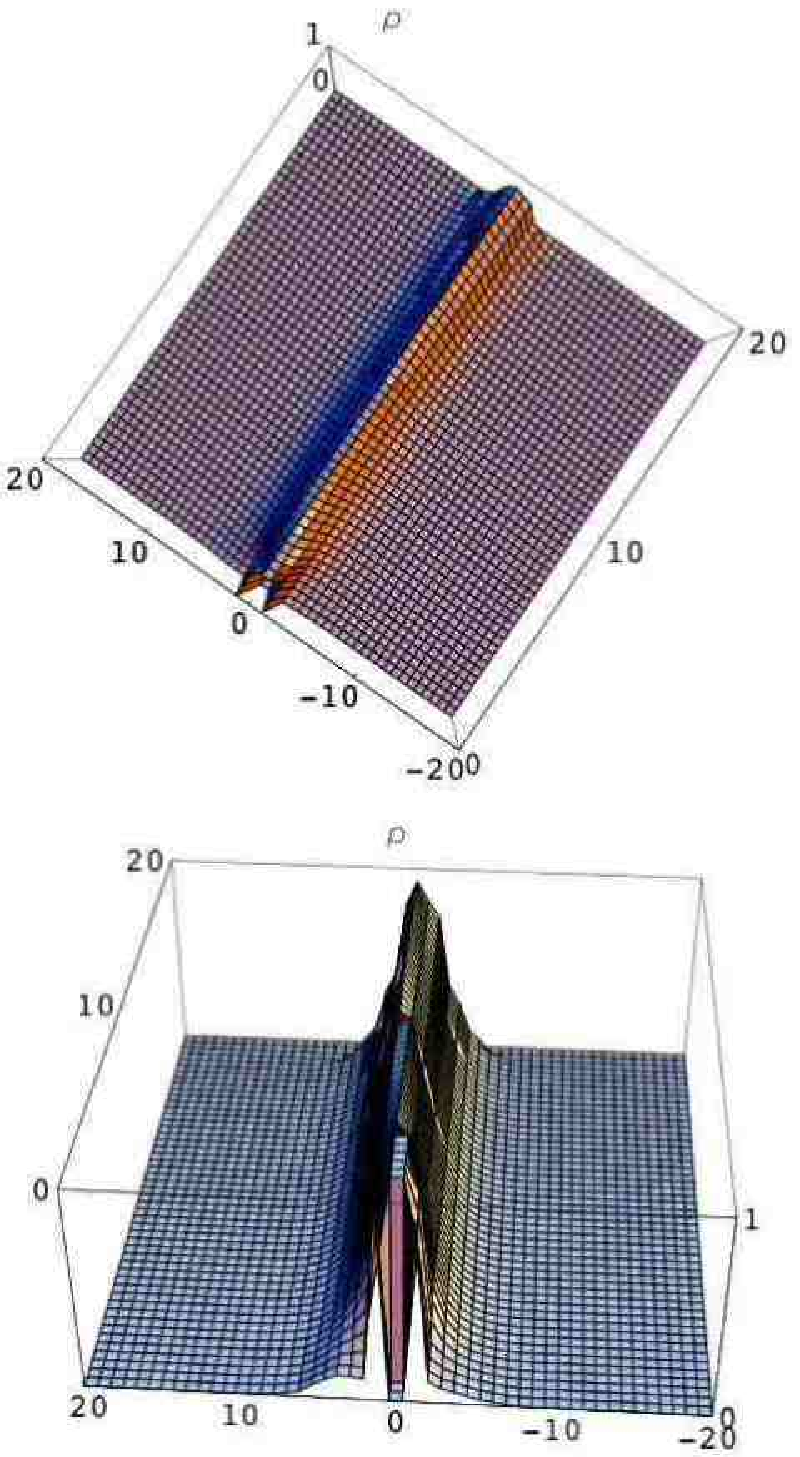}}
%\scalebox{0.5} {\includegraphics[]{edh6.eps}}
\scalebox{0.8} {\includegraphics[]{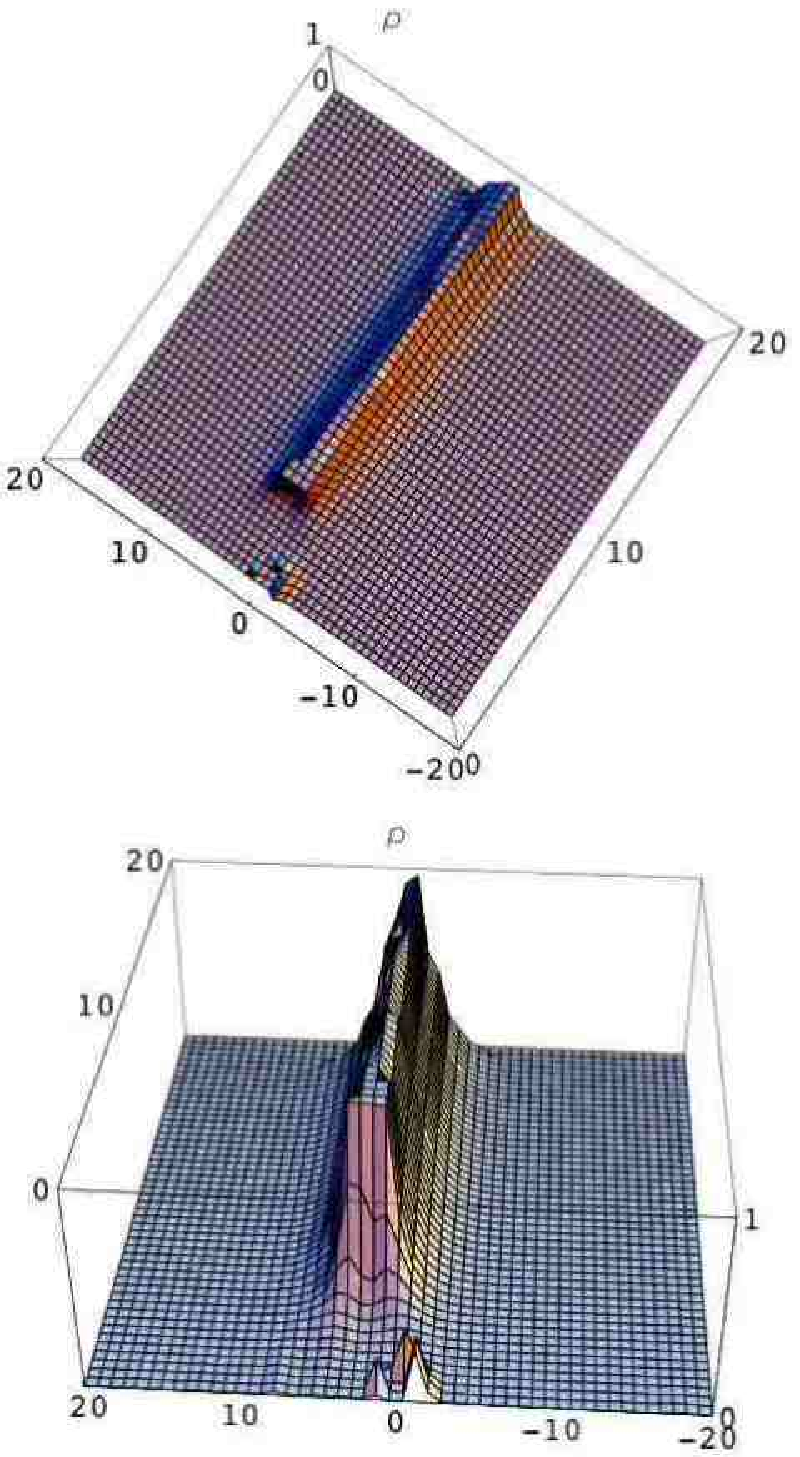}}
\scalebox{0.8} {\includegraphics[]{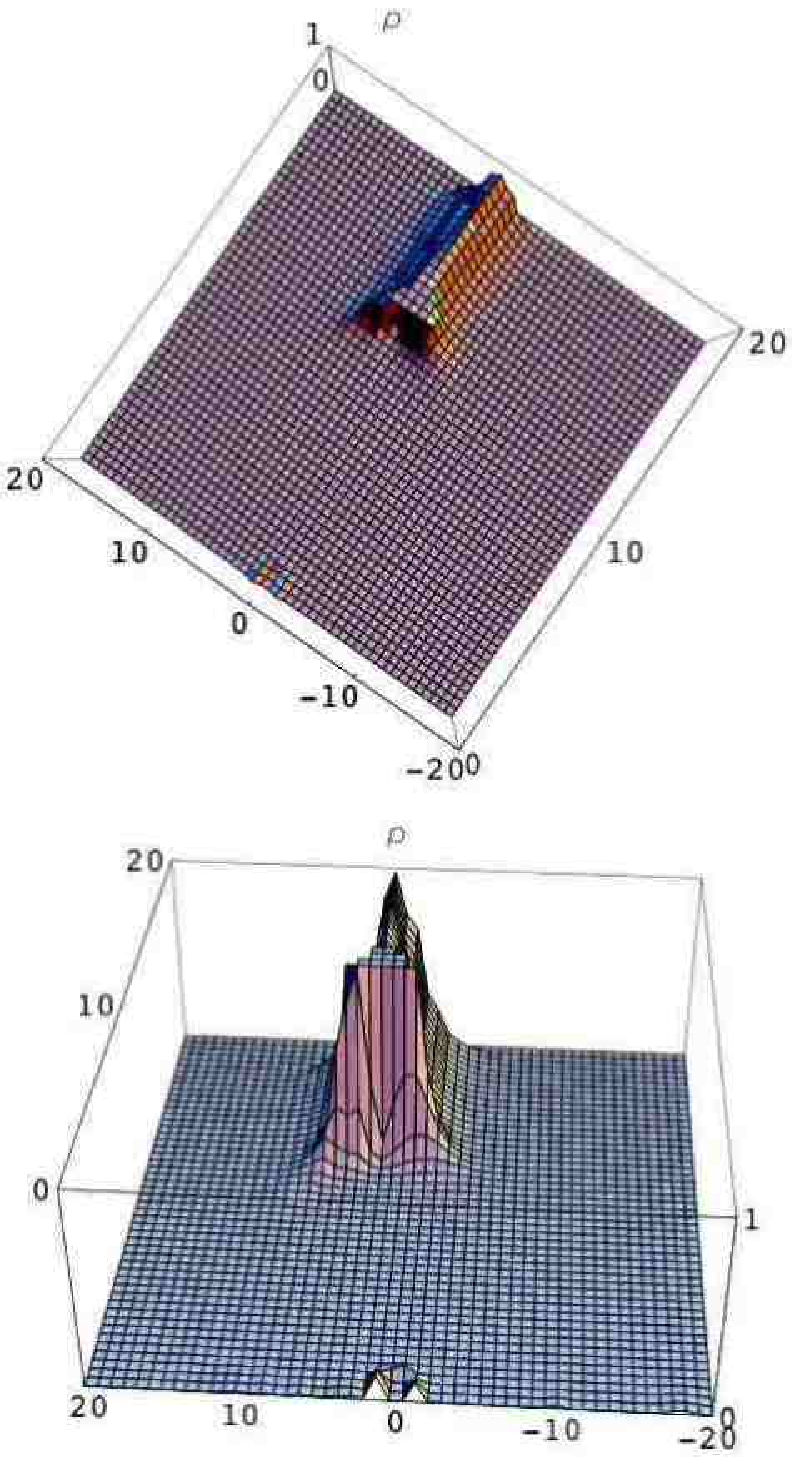}}
\caption{Snapshots of the evolution of the energy density $\rho$. The
 values used in this simulation are the ones of the previous figure. To
 emphasize the expansion of the hole, we illustrate the evolution of
 $\rho$ from two viewpoints.}
\label{fig1-2}
\end{figure*}

\begin{figure*}[th]
\scalebox{0.7} {\includegraphics[]{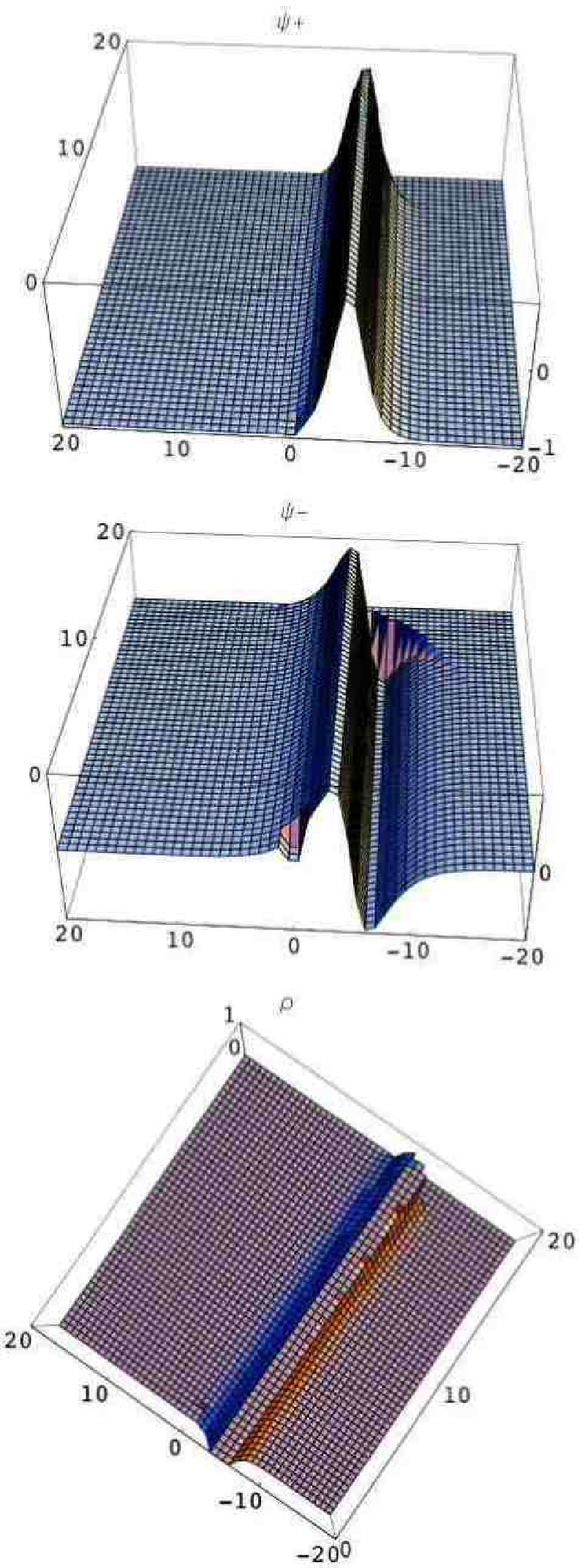}}
%\scalebox{0.7} {\includegraphics[]{pic22.eps}}
\scalebox{0.7} {\includegraphics[]{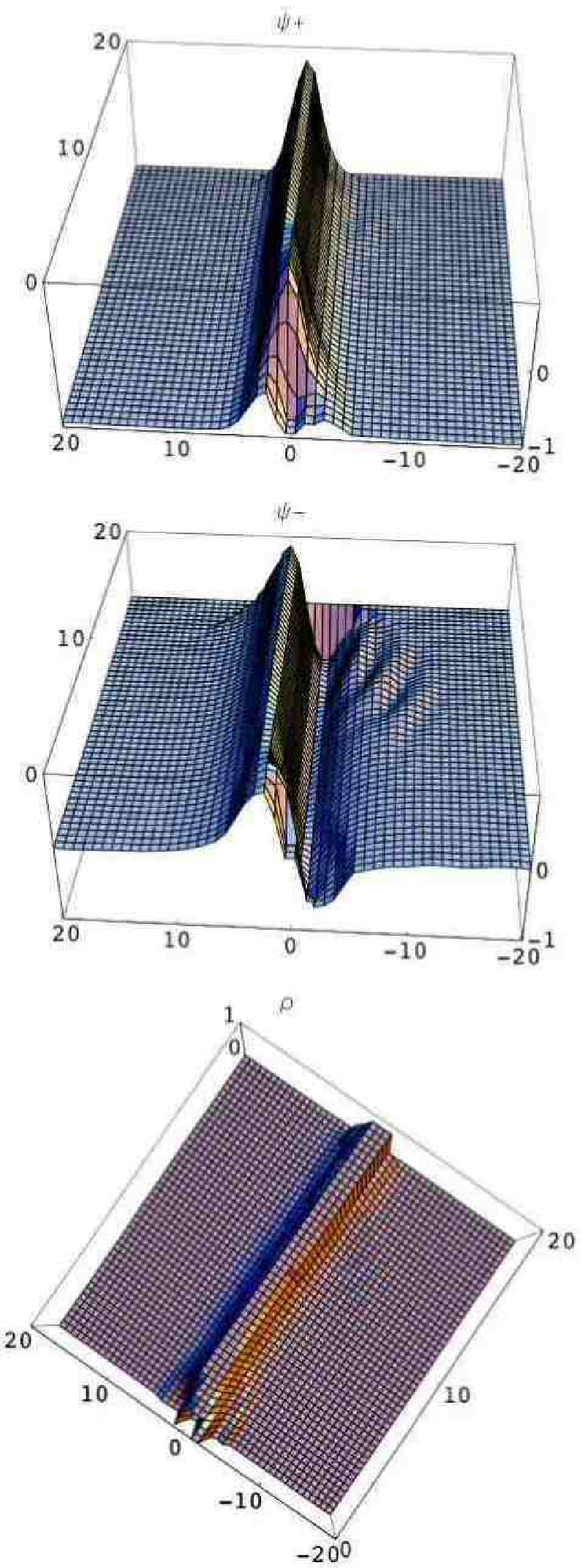}}
\scalebox{0.7} {\includegraphics[]{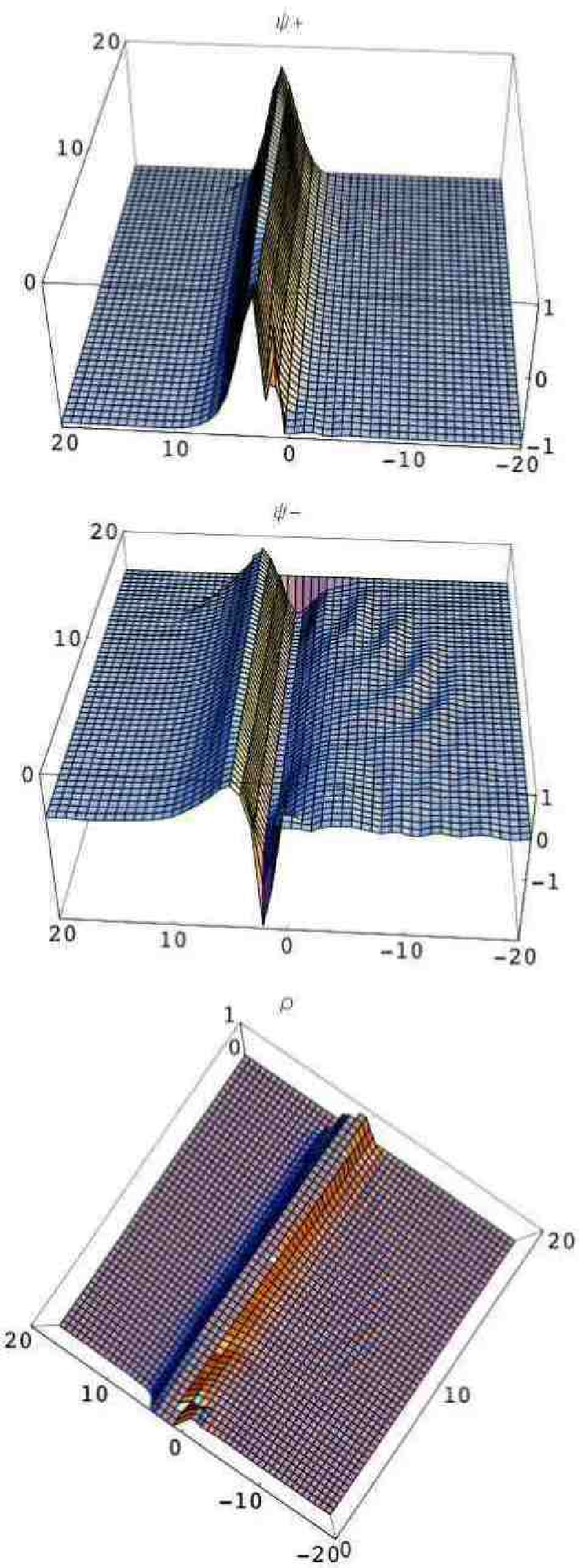}}
\caption{Snapshots of the evolution of $\psi_\pm$ and of the energy
 density $\rho$. The values used in this simulation are $\lambda=10$,
 $\mu=0.4$ and $\eta=1$. The velocity if chosen to be $0.5$. The black
 hole horizon is a circle of radius 1 centered at the 0. The spacetime
 dimensionality is 5. In this case only recombination occurs and the
 domain wall reconnects behind the black hole.}
\label{fig2-1}
\end{figure*}

\begin{figure*}[th]
\scalebox{0.8} {\includegraphics[]{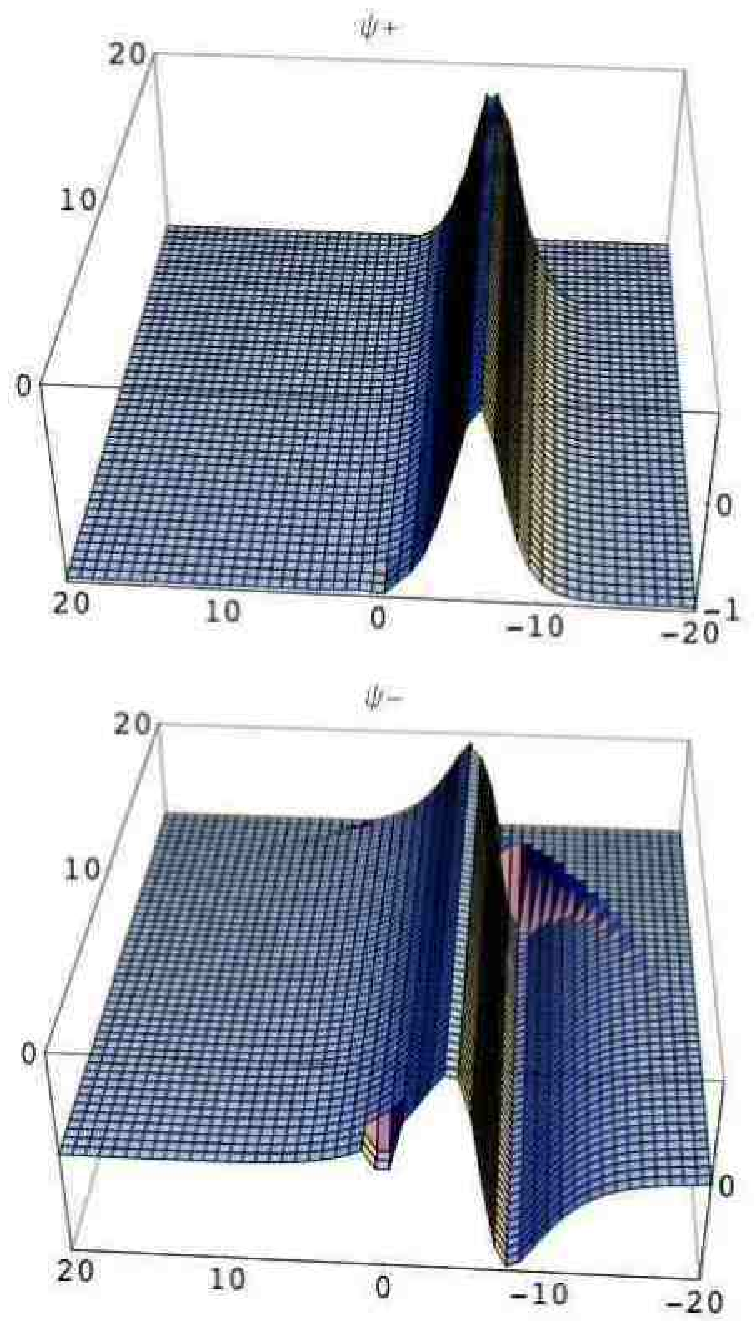}}
%\scalebox{0.6} {\includegraphics[]{f3h2.eps}}
%\scalebox{0.5} {\includegraphics[]{f3h3.eps}}
%\scalebox{0.6} {\includegraphics[]{f3h4.eps}}
\scalebox{0.8} {\includegraphics[]{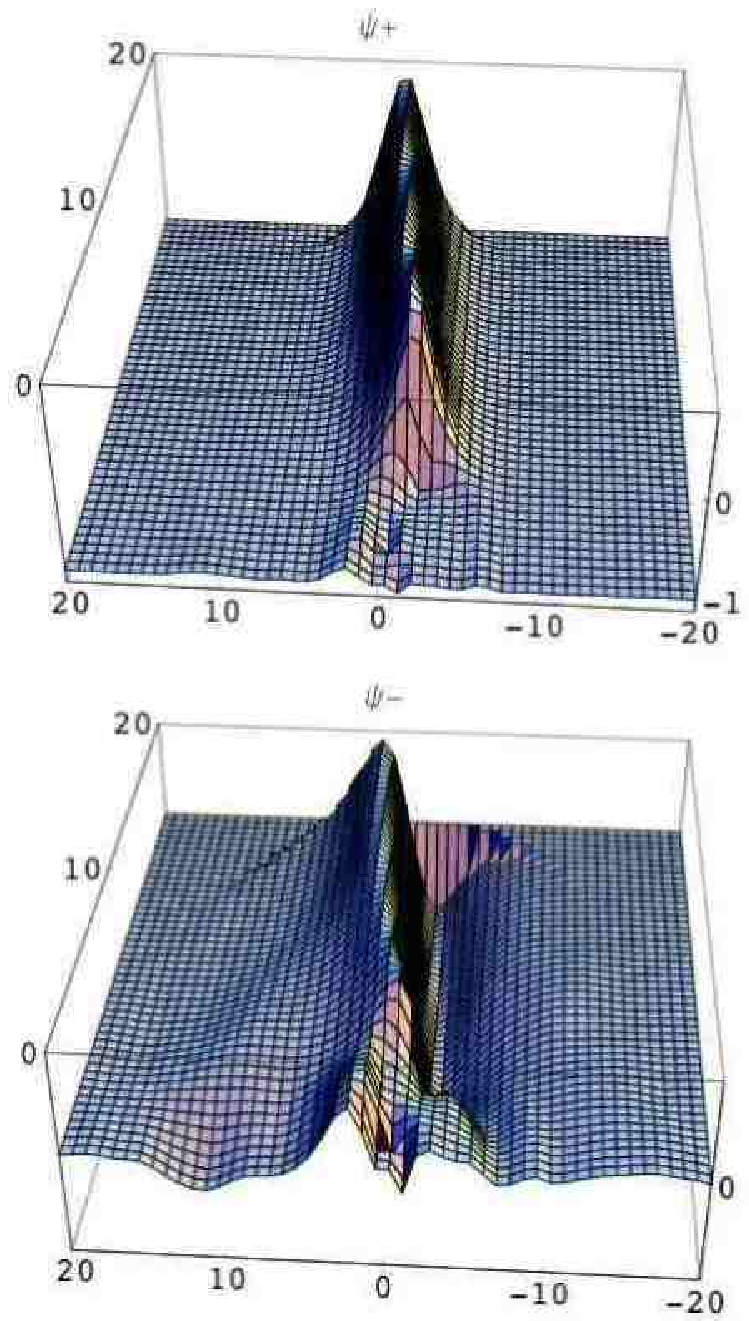}}
\scalebox{0.8} {\includegraphics[]{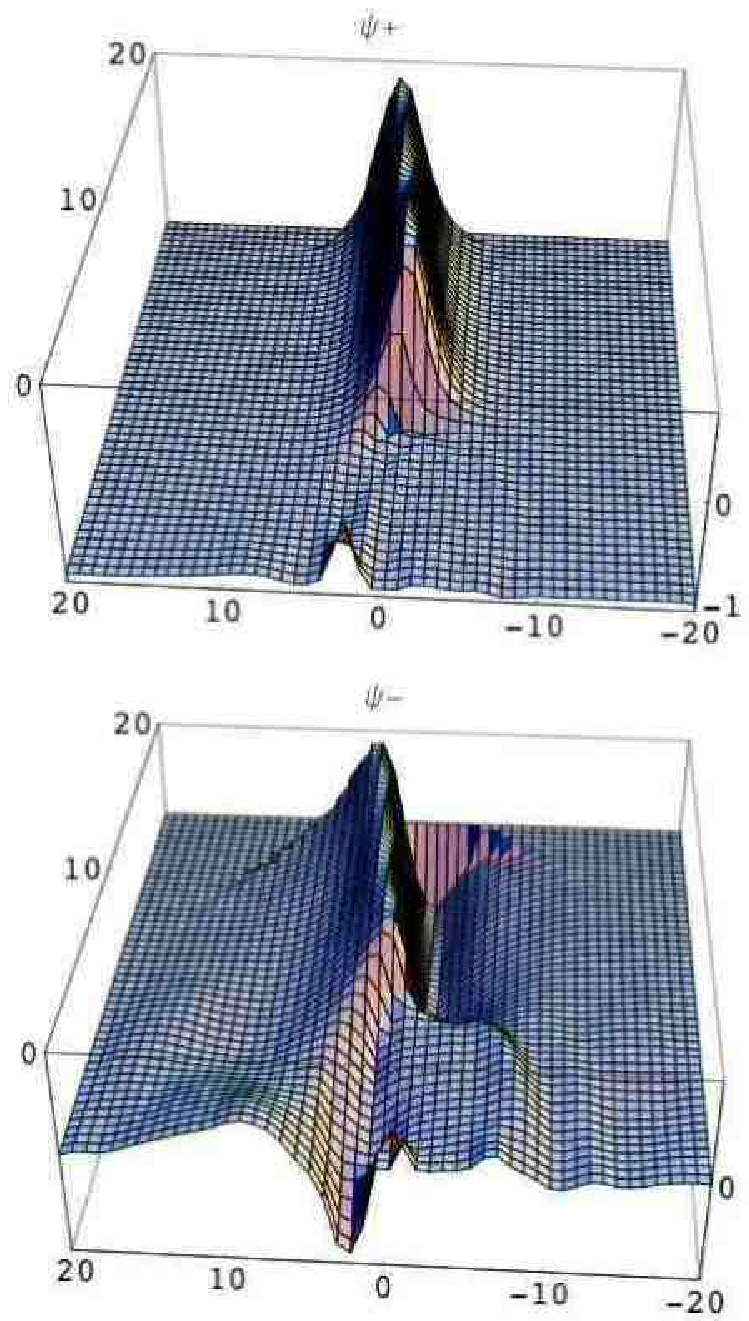}}
%\scalebox{0.5} {\includegraphics[]{f3h7.eps}}
\scalebox{0.8} {\includegraphics[]{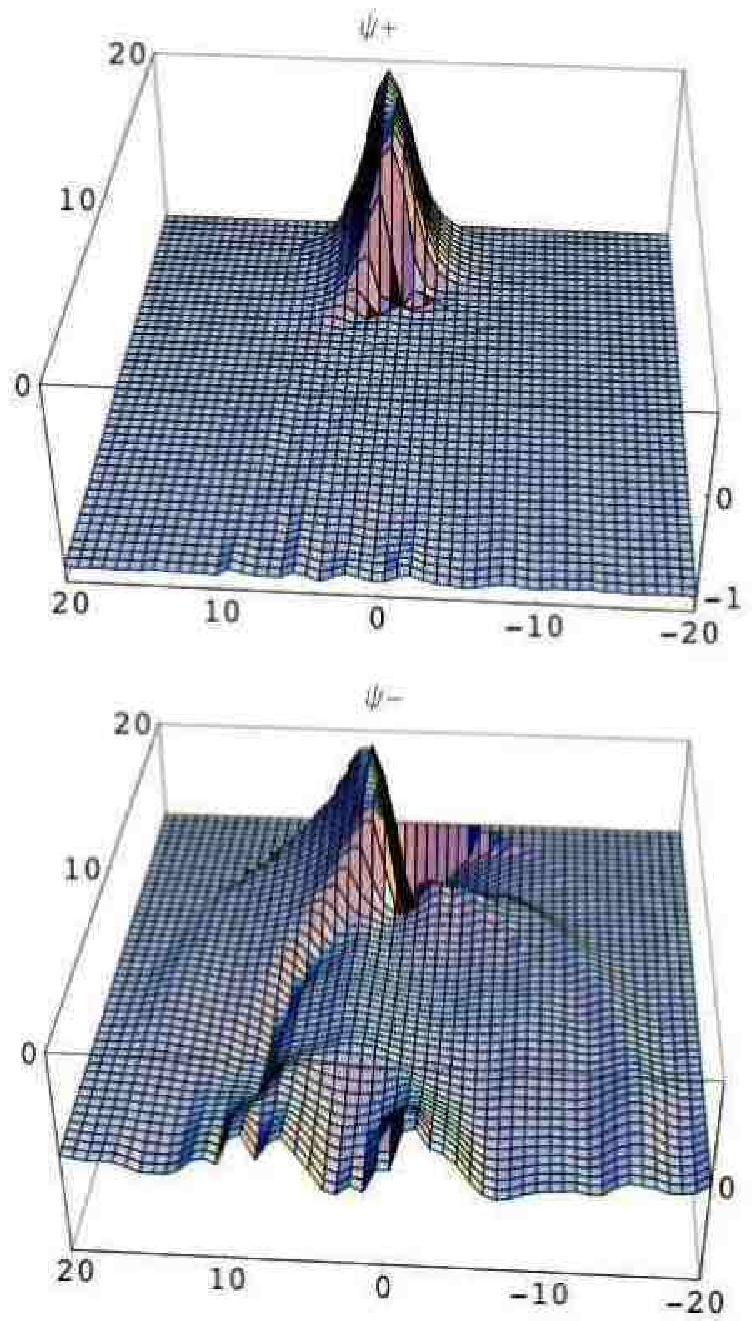}}
\caption{Snapshots of the evolution of $\psi_\pm$. The values used in
 this simulation are $\lambda=10$, $\mu=0.4$ and $\eta=1$.The velocity
 if chosen to be $0.1$. The black hole horizon is a circle of radius 1
 centered at the 0. The spacetime dimensionality is 4. One can observe
 the formation of a hole that locally destroys the wall.}
\label{fig3-1}
\end{figure*}

\begin{figure*}[th]
\scalebox{0.8} {\includegraphics[]{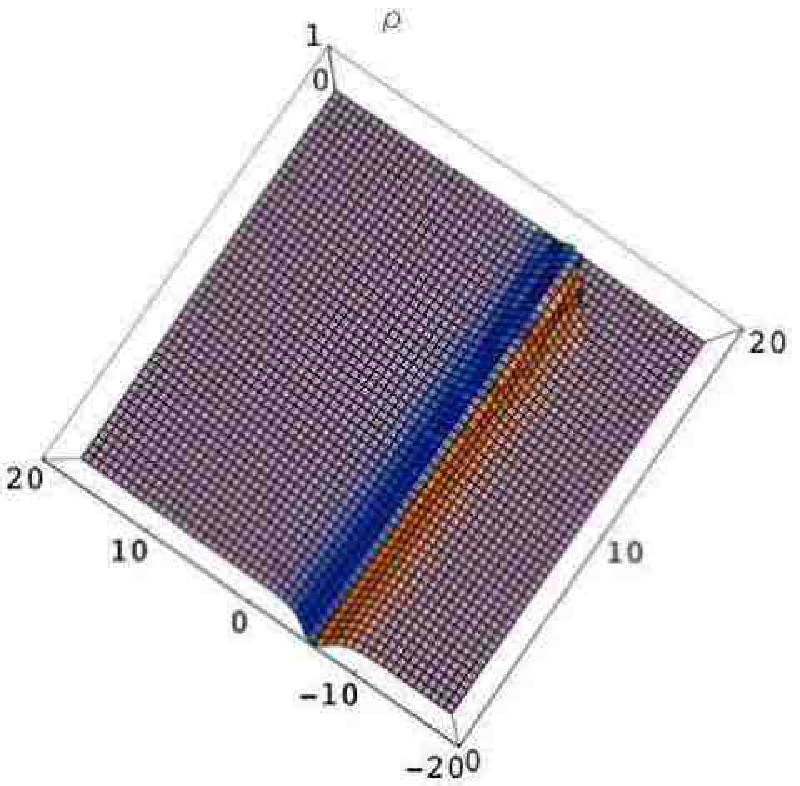}}
%\scalebox{0.6} {\includegraphics[]{ed3h2.eps}}
%\scalebox{0.6} {\includegraphics[]{ed3h3.eps}}
%\scalebox{0.5} {\includegraphics[]{ed3h4.eps}}
\scalebox{0.8} {\includegraphics[]{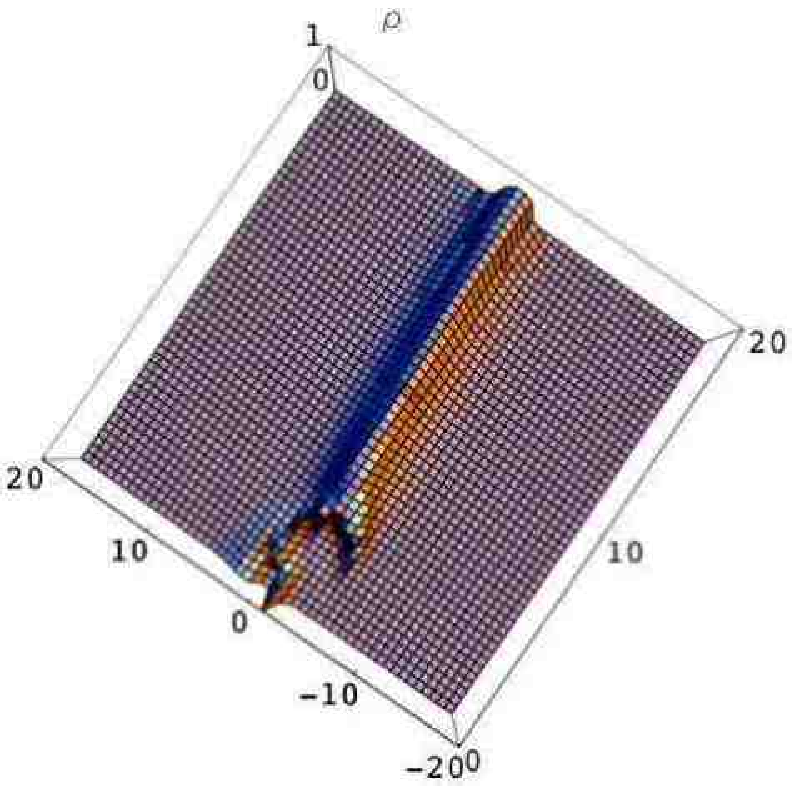}}
\scalebox{0.8} {\includegraphics[]{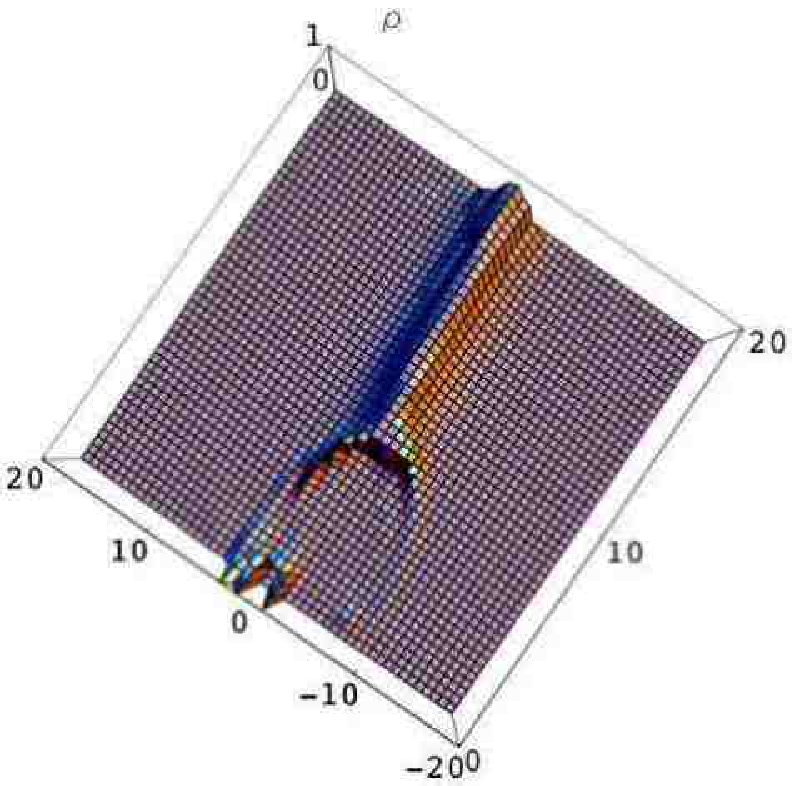}}
\scalebox{0.8} {\includegraphics[]{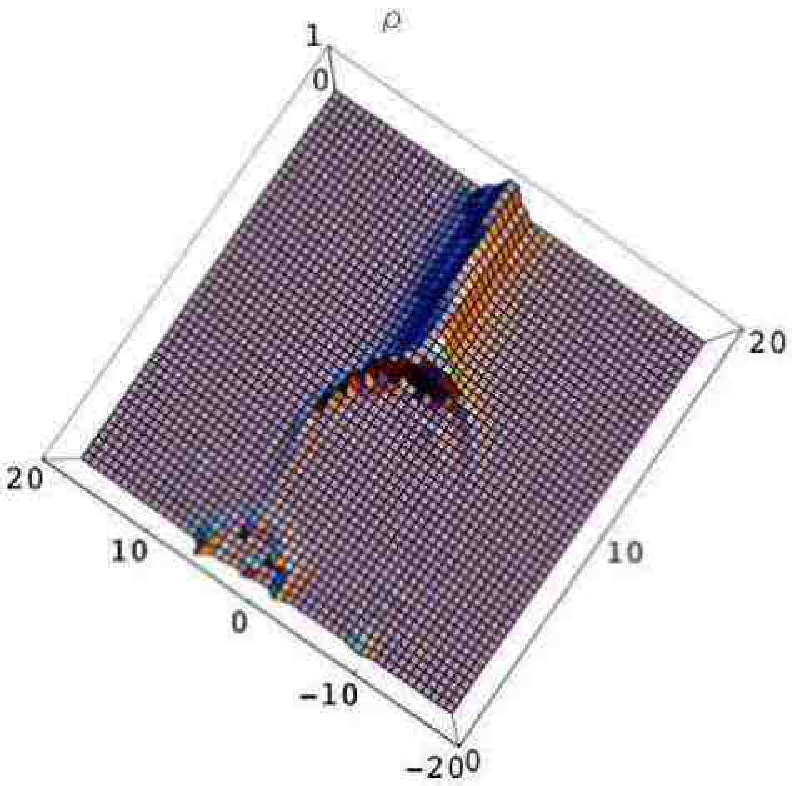}}
%\scalebox{0.5} {\includegraphics[]{ed3h8.eps}}
\caption{Snapshots of the evolution of the energy density $\rho$. The
 values used in this simulation are the ones of the previous figure.}
\label{fig3-2}
\end{figure*}

Focusing on whether a hole is formed or not, 
the results of simulations for various choices of parameters 
are summarized in Fig.~\ref{crit}.  
The dots in Fig.~\ref{crit} are calculated performing the simulation and increasing the collisional velocity at steps of $0.05$. The critical value is chosen as the mean point between the last value for which a hole forms and the first for which it does not. The error made in this estimate is taken care of by drawing the dots with a radius of $0.025$.
The plots are made for three different values of $\mu$; 
$\mu=0.2, 0.4$ and $0.8$. 
Interestingly, and contrary to our intuition, in a certain region of $\lambda$, we record
the presence of a critical velocity, above which the formation of a hole
is not observed. 
Quadratic best fit curve for the critical velocity 
is also shown for each value of $\mu$.  
From this plot we can
distinguish the parameter space into three regions, as schematically 
described in Fig.~\ref{crit_sch}; (1) domain wall solutions do 
not exist even in the flat background, (2) the domain wall is pierced by 
a black hole after collision, and (3) the domain wall is not pierced. 
Above the `critical line' the domain wall is
stable in the sense that the black hole is unable to
pierce the domain wall.

The low velocity limit in this plot will correspond to the adiabatic 
case that we have discussed above. Hence, $\lambda_2$ can be 
read from this limit of the curves for the critical velocity. 
When we prepare the initial configurations 
for the present simulations, the threshold values $\lambda_1$,  
at which a domain wall solution marginally exists 
in the flat background, are obtained simultaneously. 
Thus the obtained numerical values of $\lambda_1/\mu^2$ are 
$36.25,~43.75,~45.31$ for $\mu=0.2, 0.4$ and $0.8$, respectively. 
These values are almost the same as the values  
read from the large velocity limit of the curves 
for the critical velocity. 
For $\mu=0.2$, $r_0$ is sufficiently large and the ratio 
$\lambda_2/\lambda_1$ is close to unity.
While, for $\mu=0.8$, $r_0$ is close to unity and the ratio 
$\lambda_2/\lambda_1$ is close to 2.
These results perfectly agree with the picture presented above. 
We have performed the same check
in four and six dimensions, finding good agreement.

We briefly discuss the potential relevance for the
cosmological domain wall problem.
First of all, we notice that, contrary to our intuition, the collision with a black hole can be an
efficient way of getting rid of the domain walls in the region of
parameters where the energy scale for the string is very close 
to that for the domain wall. This is an unexpected constraint 
found in the present paper. One may think that in a concrete model, the mechanism may compete with other processes in erasing the wall. 
However, it is easy to see that this is not the case.
%In the case of the `invisible' axion model $\eta$ is of order $10^{9}- 10^{12}$ GeV, and the mass scale of the wall is of order $1$ GeV, thus much smaller than $\eta$
In fact, it can be argued that generically the perforation mechanism is not a likely solution anyway. Cosmologically $N=1$ axion-domain walls are harmless if they are associated with many strings. When two energy scales for the string and the domain wall are very close, it is almost impossible to imagine that only domain walls are formed without forming strings. In simple words, the barrier for the string formation is not very large compared to the energy `stored' in the walls, and it is reasonable to believe that any dynamical process like wall-wall collisions or fluctuations of the wall will lead to the creation of a hole in the wall. This suggests that the collision with the black holes will not play a major role in the fate of such domain walls, although it may produce minor modifications in the evolution of a domain wall - string system.

\section{conclusions}

Since it was shown that gravity can be localized on a
$(p+1)-$dimensional submanifold of a higher dimensional spacetime, the
idea that `our' universe could be a $3-$brane immersed in a higher
dimensional bulk space has been a subject of intensive study.

In this context, it is interesting, not only from an academic
perspective, to ask what happens when a collision with a brane occurs. One
may imagine that the collision occurs between two branes, as in
some stringy cosmological models, or between a brane and a
black hole, as it could happen, when considering primordial black
holes. This last problem happens to be relevant also in four dimensional
standard cosmology, when the brane is interpreted as a cosmological
domain wall.

The aim of this paper was to study such a problem, namely the collision
between a brane and a black hole in a topologically non-trivial
model. The meaning of `topologically non-trivial' is that processes,
different from recombination, can, {\it in principle}, occur. Specifically we
concentrate our attention to the possibility that the brane is
perforated due to the collision. This can occur when the eventual hole
is bounded by a string, and, as working model, we consider axion type
effective field theories, which account for hybrid defects (domain walls
attached to strings), that, in our treatment, will model the brane.

The problem of the collision is investigated by numerically solving the
Klein-Gordon equations for the brane in the black hole
spacetime. We have scanned the process by varying the parameters of the
potential, $\mu$ and $\lambda$, and this also effectively takes account of varying the
size of the black hole, due to the scaling properties of the Lagrangian.

The results of the simulation are characterized, for fixed $\mu$, 
by two critical values of the parameter $\lambda$: $\lambda_1$ and
$\lambda_2$. The first one gives a lower bound on the existence of a domain
wall solution, whereas, the other value, $\lambda_2$, marks the threshold above which the domain
wall is stable in the sense that cannot be perforated by the black hole. 
The critical value $\lambda_2$ separates the case where an unstable static domain wall configuration with 
a hole edged by a sting on the equatorial plane of the black hole 
exists and the case in which such a configuration does not.  
When the string thickness $r_s$ exceeds 
$\Delta$, the proper distance 
between the horizon and the string core, 
the unstable configuration does not exist. 
The hole cannot shrink and the domain wall is necessarily pierced.
On the other hand, for $r_s<\Delta$ a static configuration of a domain wall with
a floating string exists. Consequently, the hole when created, is accreted by the black hole. 
As a result the wall is not pierced. 

In the intermediate region between $\lambda_1$ and $\lambda_2$, 
where the domain wall can be perforated, we observe the existence of a critical collisional velocity,
$v_{crit}$, (unexpectedly) above which the domain wall is also
stable. This fact is illustrated with the aid of the phase diagram of Fig.~\ref{crit}. 
We have numerically evaluated the ratio 
$\lambda_2/\lambda_1$, and compared it with the one estimated from
analytic arguments using the static configuration, finding agreement.

Finally, we have discussed the relevance of the above process in the
context of a recently conjectured solution to the cosmological domain
wall problem. We argue that the
mechanism is efficient only when domain walls form 
immediately after strings. 
However, in such cases, string formation cannot be suppressed when 
domain walls are formed, because the potential barrier to overcome in order to have string production is comparable to the energy stored in the domain walls. 
Hence, those domain walls are quickly eaten up by the expanding holes (formed before the collisions) and disappear 
anyway. Therefore it seems unlikely that this mechanism 
plays an important role as a solution to the the cosmological 
domain wall problem, even though may have a small effect in the evolution of domain wall networks.

\acknowledgements
AF is supported by the 21st Century COE ``Center for Diversity and
Universality in Physics'' at Kyoto university, and by the
Yukawa Memorial Foundation. TT is supported by Grant-in-Aid for Scientific
Research, No. 16740141 and by Monbukagakusho Grant-in-Aid
for Scientific Research(B) No.~17340075. 
This work is also supported in part by the 21st Century COE ``Center for
Diversity and Universality in Physics'' at Kyoto university, from the Ministry of
Education, Culture, Sports, Science and Technology of Japan
and also by the Japan-U.K. Research Cooperative Program
both from Japan Society for Promotion of Science.

\newpage

%5D hole %

%5D no hole %

%4D hole %

\end{document}